\documentclass[epj]{svjour}
\usepackage{graphicx,amsmath}
\usepackage[T1]{fontenc}
\usepackage[latin1]{inputenc}
\newcommand{\figwidth}{1.0\columnwidth}
\newcommand{\eq}[1]{Eq.(\ref{#1})}
\newcommand{\fig}[1]{Fig.~\ref{#1}}

\newcommand{\avg}[1]{ {\langle #1 \rangle} }

\newcommand{\ahum}[1]{``#1''}

\begin{document}

\title{The isotropic-to-nematic transition in a two-dimensional fluid of hard 
needles : a finite-size scaling study}

\titlerunning{IN transition in 2D needles}

\author{R. L. C. Vink}
\authorrunning{RLC Vink}

\institute{Institute of Theoretical Physics \\ Georg-August-Universit\"at 
G\"ottingen \\ Friedrich-Hund-Platz~1 \\ 37077 G\"ottingen, Germany}

\date{Received: \today / Revised version: xx-xx-xx}

\abstract{The isotropic-to-nematic transition in a two-dimensional fluid of hard 
needles is studied using grand canonical Monte Carlo simulations, multiple 
histogram reweighting, and finite size scaling. The transition is shown to be of 
the Kosterlitz-Thouless type, via a direct measurement of the critical exponents 
$\eta$ and $\beta$, of the susceptibility and order parameter, respectively. At 
the transition, $\eta=1/4$ and $\beta=1/8$ are observed, in excellent agreement 
with Kosterlitz-Thouless theory. Also the shift in the chemical potential of the 
nematic susceptibility maximum with system size is in good agreement with 
theoretical expectations. Some evidence of singular behavior in the density 
fluctuations is observed, but no divergence, consistent with a negative specific 
heat critical exponent. At the transition, a scaling analysis assuming a 
conventional critical point also gives reasonable results. However, the apparent 
critical exponent $\beta_{\rm eff}$ obtained in this case is not consistent with 
theoretical predictions.
}

\maketitle

\section{Introduction} 

Upon increasing density, a fluid of hard needles in two dimensions 
undergoes a transition from an isotropic to a nematic phase 
\cite{physreva.31.1776}. In the isotropic phase, the orientational 
correlations decay exponentially to zero, while in the nematic phase 
algebraic decay is observed \footnote{For this reason, the nematic phase 
should perhaps be termed quasi-nematic. For notational convenience, 
however, we refrain from doing so in this paper.}. In the thermodynamic 
limit, long-range nematic order is thus absent in both phases, and the 
available evidence points to a transition of the Kosterlitz-Thouless 
(KT) type \cite{physreva.31.1776,citeulike:4464987}. In other words, 
the universality class of the isotropic-to-nematic (IN) transition in 
two-dimensional hard needles should be that of the XY model 
\cite{kosterlitz:1974}, and one expects to find the same set of critical 
exponents. For the XY model, the latter are known exactly, but their 
verification in a fluid of hard needles remains elusive to this day. The 
purpose of this paper is to fill this gap, using grand canonical Monte 
Carlo simulations and finite-size scaling. Indeed, our simulations 
consistently recover the XY exponents $\eta=1/4$ and $\beta=1/8$, of the 
susceptibility and order parameter, respectively. In addition, the 
scaling of the chemical potential at the susceptibility maximum is in 
good agreement with XY universality. Hence, our data quantitatively 
confirm the KT scenario in fluids of hard needles.

We also observe that, at high density in the nematic phase, the decay of 
nematic order with increasing system size is very slow. This means that 
even in macroscopic samples a substantial degree of nematic order is 
present. The same occurs in the XY model: even though long-range 
magnetic order is absent in the thermodynamic limit, finite XY systems 
at low temperature nevertheless reveal considerable magnetic order. The 
consequences of this have been worked out by Bramwell and Holdsworth 
(BH), who conclude that the formation of magnetic order in finite XY 
models is characterized by an effective critical exponent $\beta_{\rm 
eff}$ \cite{bramwell.holdsworth:1993}. Interestingly, when we analyze 
our data assuming a conventional critical point, we can also 
consistently measure such effective exponents, which moreover obey the 
hyperscaling relation. Hence, the BH scenario for the XY model seems to 
be valid in fluids of hard needles also, even though $\beta_{\rm eff}$ 
obtained by us differs from the XY value predicted by BH.

The outline of this paper is as follows. We first specify the model 
details and the simulation method. Next, we present our raw simulation 
data, displaying how the various observables of interest depend on the 
chemical potential and system size. The raw data is then analyzed using 
several finite size scaling methods. We end with a discussion and 
summary.

\section{Model and simulation method}

We consider a two-dimensional system of infinitely thin rods of length $l$, 
henceforth referred to as needles. We emphasize that our model is not 
discretized in any way: both the needle positions and orientations are 
continuous. In what follows, $l$ will be the unit of length. The needles are 
hard, i.e.~they are not allowed to overlap, and trivial factors of inverse 
temperature are set to unity throughout. The simulations are performed in the 
grand canonical ensemble, i.e.~at constant chemical potential $\mu$ and system 
area $A$, while the number of needles $N$ fluctuates. The average needle density 
increases with $\mu$ and this can be used to induce the IN transition. Hence, 
$\mu$ is the control parameter, analogous to inverse temperature in thermotropic 
systems. We use a two-dimensional simulation square of size $A=L^2$ with 
periodic boundary conditions. Insertion and removal of needles are attempted 
with equal probability, and accepted with the standard grand canonical 
Metropolis probabilities \cite{landau.binder:2000,frenkel.smit:2001}. During 
insertion, a random location in the system is selected and a needle with 
randomly selected orientation is tentatively placed at this location. If this 
needle overlaps with any of the other needles already present, the move is 
rejected. Otherwise, the new state is accepted with probability
\begin{equation}
 A(N \to N+1) = \min \left[1 , \frac{A e^\mu}{N+1} \right],
\end{equation}
with $N$ being the number of needles in the system at the start of the move. 
Similarly, during removal, one of the needles is selected at random and deleted 
from the system, and the resulting state is accepted with probability
\begin{equation}
 A(N \to N-1) = \min \left[1 , \frac{N }{A e^\mu} \right].
\end{equation}
To facilitate the efficient detection of overlap during particle insertion, a 
link-cell neighbor list is used \cite{allen.tildesley:1989}. The simulation data 
are collected as two-dimensional histograms $H_{\mu,L}(S,N)$, counting how often 
a state with nematic order parameter $S$ and particle number $N$ is observed 
(note the dependence on $\mu$ and $L$). For system sizes $L = 10 - 30$, 
histograms are obtained for several values of $\mu$; the multiple histogram 
method \cite{ferrenberg.swendsen:1989,newman.barkema:1999} is used to evaluate 
properties at intermediate values. The nematic order parameter $S$ is defined as 
the maximum eigenvalue of the orientational tensor $Q_{\alpha\beta} = 
\sum_{i=1}^N \left( 2 d_{i\alpha} d_{i\beta} - \delta_{\alpha\beta} \right)$, 
with $d_{i\alpha}$ the $\alpha$ component ($\alpha = x,y$) of the orientation 
$\vec{d}_i$ of molecule $i$, $|\vec{d}_i|=1$, and $\delta_{\alpha\beta}$ the 
Kronecker delta. We emphasize that the nematic order parameter $S$ defined in 
this way is an {\it extensive} quantity. In cases where the number of particles 
$N$ is constant, it is convenient to use the normalized {\it intensive} 
definition $S^\star = S/N$, since then one has $S^\star=0$ and $S^\star=1$, in 
an isotropic and perfectly aligned sample, respectively. However, in the grand 
canonical ensemble, $N$ is a fluctuating quantity, which itself might exhibit 
singular behavior, and so this convention is not used here.

Most of the simulations were performed on Intel DualCore processors clocked at 2 
GHz. For each system size $L$, around 10~histograms were collected, with $\mu$ 
taken from the range $\sim 5.0 - 5.2$. At these values, the needle density $\rho 
\sim 7$, which is close to the transition density observed in previous studies 
\cite{physreva.31.1776,citeulike:4464987}. Each histogram was simulated for 
$\approx 10^5$ grand canonical sweeps \cite{vink:2006}, with a sweep being 
defined as one complete renewal of the particle population (recall that the 
number of particles fluctuates). The computational effort per sweep depends on 
$\mu$ and $L$. For $\mu=5.1$ and $L=10$, $n \approx 2.3 \times 10^5$ grand 
canonical Monte Carlo attempts are required to complete one sweep; for $L=30$, 
this increases to $n \approx 2.7 \times 10^6$. The simulations began with empty 
boxes, and the first 1000~sweeps were discarded for equilibration.

\section{Results}

\subsection{Observables}

The observables of interest are the average needle density and 
the compressibility
\begin{equation}
 \rho = \avg{N}/A, \hspace{5mm} 
 \chi_\rho = \left( \avg{N^2} - \avg{N}^2 \right) / A,
\end{equation}
the nematic density (order parameter) and the nematic susceptibility
\begin{equation}\label{eq:sus}
 \sigma = \avg{S}/A, \hspace{5mm}
 \chi_\sigma = \left( \avg{S^2} - \avg{S}^2 \right) / A,
\end{equation}
and the Binder cumulant
\begin{equation}
 U_4 = \avg{S^2}^2 / \avg{S^4}.
\end{equation}
The above quantities will generally depend on $\mu$ and $L$, especially 
in the vicinity of phase transitions.

\subsection{Raw simulation data}

\begin{figure}
\begin{center}
\includegraphics[clip=,width=\figwidth]{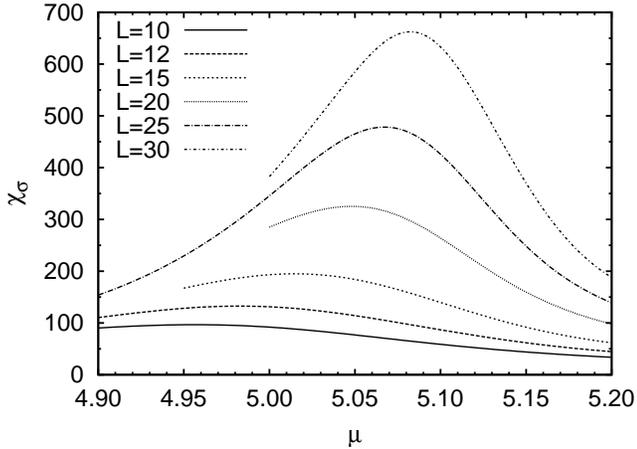}
\caption{\label{sus_raw} Plot of the nematic susceptibility $\chi_\sigma$ versus 
the chemical potential $\mu$, for several system sizes $L$ as indicated. Clearly 
visible is that $\chi_\sigma$ attains a maximum, and that the maximum grows with 
increasing system size. Note also that the position of the maximum is size 
dependent.}
\end{center}
\end{figure}

\begin{figure}
\begin{center}
\includegraphics[clip=,width=\figwidth]{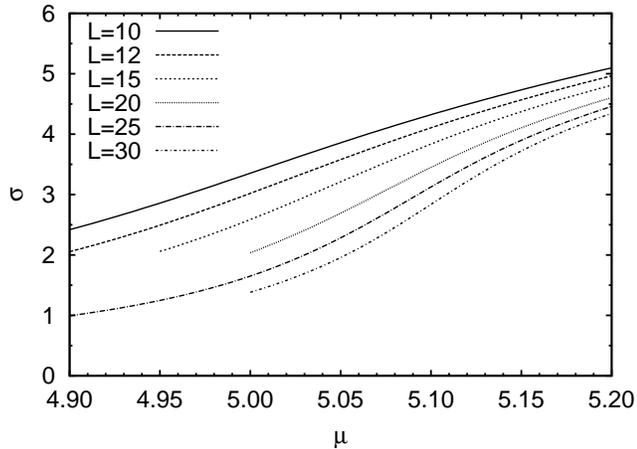}
\caption{\label{op_raw} Plot of the nematic density $\sigma$, which plays the 
role of the order parameter, versus the chemical potential $\mu$, for several 
system sizes $L$ as indicated. Note that $\sigma$ increases with $\mu$, but also 
that it overall decreases with increasing $L$. This result is compatible with 
the absence of nematic order in the thermodynamic limit.}
\end{center}
\end{figure}

\begin{figure}
\begin{center}
\includegraphics[clip=,width=\figwidth]{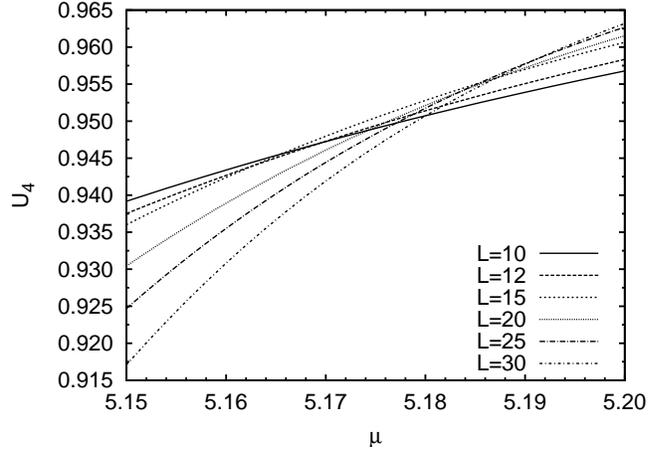}
\caption{\label{u4} Plot of the Binder cumulant $U_4$ versus the chemical 
potential $\mu$, for several system sizes $L$ as indicated. Note that the data 
from the various system sizes approximately intersect. For increasing $L$, a 
shift of the intersection point toward higher values of $U_4$ is visible.}
\end{center}
\end{figure}

\begin{figure}
\begin{center}
\includegraphics[clip=,width=\figwidth]{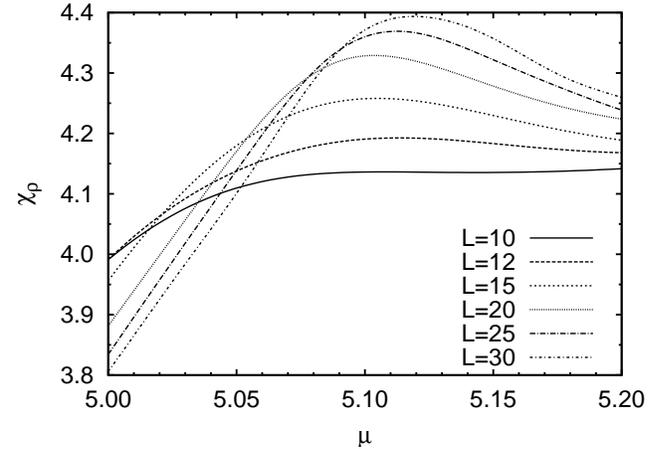}
\caption{\label{cv_raw} Plot of the density fluctuation (compressibility) 
$\chi_\rho$ versus the chemical potential $\mu$, for several system sizes $L$ as 
indicated. Note the presence of the maximum, but also that the increase of the 
maximum with $L$ is much milder compared to that of $\chi_\sigma$ in 
\fig{sus_raw}.}
\end{center}
\end{figure}

\begin{figure}
\begin{center}
\includegraphics[clip=,width=\figwidth]{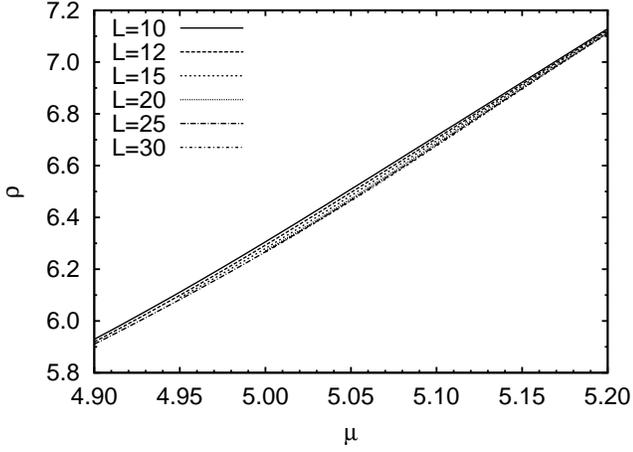}
\caption{\label{rho} Plot of the particle density $\rho$ versus the chemical 
potential $\mu$, for several system sizes $L$ as indicated. The overall trend is 
that $\rho$ increases with $\mu$, and that it decreases mildly with $L$.}
\end{center}
\end{figure}

\begin{figure}
\begin{center}
\includegraphics[clip=,width=\figwidth]{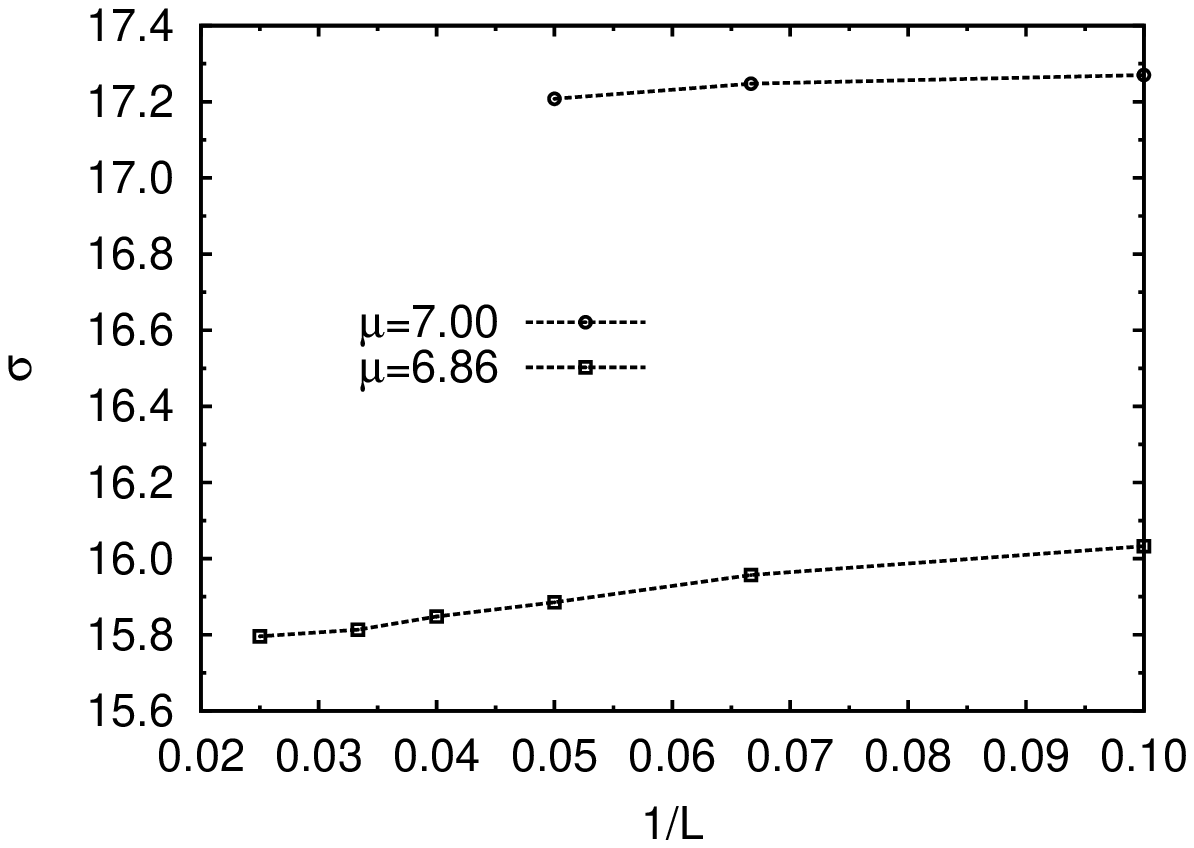}
\caption{\label{decay} Variation of the nematic density $\sigma$ (order 
parameter) versus $1/L$ measured at two high values of the chemical potential 
$\mu$ as indicated (by high is meant well beyond the maxima in $\chi_\sigma$ and 
$\chi_\rho$). The important result to take from this graph is that $\sigma$ does 
not saturate at a finite value, but continues to decrease with increasing $L$, 
consistent with the absence of true nematic order in the thermodynamic limit.}
\end{center}
\end{figure}

We first present our raw simulation data. In \fig{sus_raw} the nematic 
susceptibility $\chi_\sigma$ is plotted versus $\mu$ for several system 
sizes. The nematic susceptibility displays a maximum, which becomes more 
pronounced as $L$ increases. In addition, the chemical potential at the 
maximum $\mu_L^\star$ depends on $L$. The nematic density is shown in 
\fig{op_raw}. We observe that $\sigma$ increases monotonically with 
$\mu$, while it decreases with increasing $L$. The Binder cumulant $U_4$ 
is shown in \fig{u4}. The data from the different system sizes 
approximately intersect. The compressibility $\chi_\rho$ is shown in 
\fig{cv_raw}. As with $\chi_\sigma$, the formation of a maximum is 
visible, but it increases only mildly with $L$. In \fig{rho} the needle 
density is shown. We observe a monotonic increase of $\rho$ with $\mu$, 
and a weak decrease of $\rho$ with $L$. In \fig{decay}, we show the 
nematic density at high values of the chemical potential, chosen well 
beyond the extrema of $\chi_\sigma$ and $\chi_\rho$. We observe that 
$\sigma$ does not saturate, but continues to decrease with increasing $L$.

The raw simulation data already provide evidence of a phase transition. Based on 
\fig{sus_raw}, the transition is characterized by a diverging nematic 
susceptibility. There is also evidence of singular behavior in the 
compressibility, see \fig{cv_raw}, but it is much weaker. The key observation is 
that the transition does not yield any finite order parameter: even at very high 
chemical potential, the nematic density $\sigma$ does not saturate, but 
continues to decay with increasing $L$, see \fig{decay}. This is consistent with 
previous simulations of hard needles \cite{physreva.31.1776,citeulike:4464987}, 
and provides further confirmation that nematic order is most likely absent in 
the thermodynamic limit, i.e.~$\sigma$ decays to zero as $L \to \infty$ 
irrespective of $\mu$. The absence of nematic order appears to be a general 
property of two-dimensional liquid crystals -- computer simulations of rods 
reveal similar behavior \cite{lagomarsino.dogterom.ea:2003,bates.frenkel:2000} 
-- and is conform the Mermin-Wagner theorem \cite{physrevlett.17.1133}. Note 
that for certain liquid crystal pair potentials, the absence of nematic order 
can be proved rigorously \cite{physreva.4.675}. The observation that the order 
parameter vanishes in the thermodynamic limit rules out a conventional critical 
point, leaving a transition of the KT type as the most likely alternative.

\section{Finite size scaling}

\subsection{KT scaling}

Characteristic of a KT transition is the exponential divergence of the 
correlation length \cite{kosterlitz:1974}. If one starts in the 
isotropic phase, and moves toward the nematic phase by increasing the 
chemical potential, the correlation length diverges as
\begin{equation}\label{eq:cl}
 \xi \propto \exp \left( b t^{-1/2} \right), \, 
 t \equiv \mu_\infty - \mu, \, t \geq 0,
\end{equation}
with $\mu_\infty$ the chemical potential at the transition, and 
nonuniversal constant $b>0$. As $\xi$ diverges faster than any power 
law, the conventional critical exponent $\nu$ of the correlation length 
does not exist, but it is possible to define exponents $\beta$ and 
$\eta$, of the nematic order parameter $\sigma$ and susceptibility 
$\chi_\sigma$, respectively, by expressing these quantities in terms of 
$\xi$
\begin{equation}
 \sigma \propto \xi^{-\beta}, \hspace{5mm} 
 \chi_\sigma \propto \xi^{2-\eta},
\end{equation}
with KT values $\beta=1/8$ and $\eta=1/4$ \cite{kosterlitz:1974}. We 
emphasize that these exponents are only observed on the positive 
interval $0 \leq t < \epsilon$, with $\epsilon$ not too large 
\cite{citeulike:4540692,citeulike:4540703}. In the regime $t<0$, the 
correlation length remains infinite, and the exponents become functions 
of~$t$.

\begin{figure}
\begin{center}
\includegraphics[clip=,width=\figwidth]{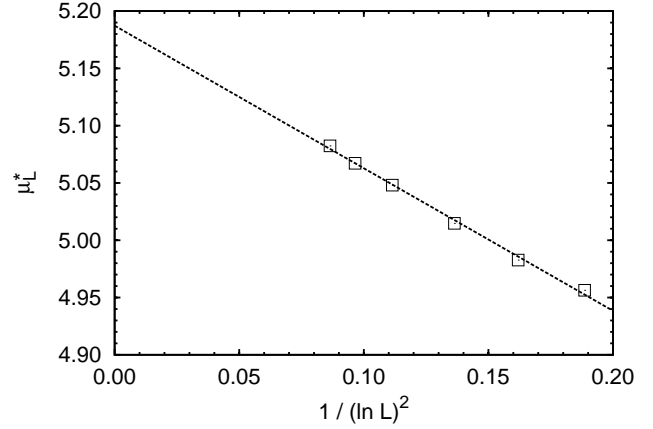}
\caption{\label{kt_fit} Determination of the thermodynamic limit transition 
chemical potential $\mu_\infty$ assuming the KT scenario. Plotted is the 
chemical potential $\mu_L^\star$ of the nematic susceptibility maximum versus 
$1/(\ln L)^2$; the line is a fit to the KT form of \eq{eq:fit}. The data follow 
the KT prediction well, and from the fit $\mu_\infty \approx 5.187$ is obtained.}
\end{center}
\end{figure}

\begin{figure}
\begin{center}
\includegraphics[clip=,width=\figwidth]{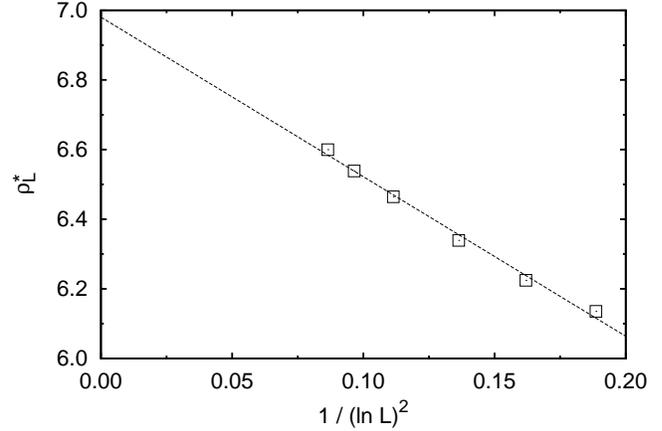}
\caption{\label{rho_kt} Determination of the thermodynamic limit transition 
density $\rho_\infty$ assuming the KT scenario. Plotted is the density 
$\rho_L^\star$ obtained at $\mu = \mu_L^\star$ versus $1/(\ln L)^2$; the line is 
a fit to \eq{eq:fit}. From the fit $\rho_\infty \approx 6.98$ is obtained.}
\end{center}
\end{figure}

In finite systems, the $L$ dependence of $\sigma$ and $\chi_\sigma$ in 
the regime $0 \leq t < \epsilon$ is described in the context of finite 
size scaling by
\begin{equation}\label{eq:ansatz}
 \sigma (L) = L^{-\beta} f_1 (L/\xi), \hspace{5mm}
 \chi_\sigma (L) = L^{2-\eta} f_2 (L/\xi),
\end{equation}
with scaling functions $f_i$. Regarding $\chi_\sigma$, this implies that 
the chemical potential $\mu_L^\star$ of the susceptibility maximum, see 
also \fig{sus_raw}, must occur at the same argument of the scaling 
function
\begin{equation}
 \left. \frac{L}{\xi} \right|_{\mu = \mu_L^\star} = c,
\end{equation}
with $c$ a constant of order unity. By substitution of \eq{eq:cl} one easily 
derives that, to leading order, $\mu_L^\star$ is shifted from $\mu_\infty$ as 
\cite{citeulike:4518230,citeulike:4478384,citeulike:4472295}
\begin{equation}\label{eq:fit}
 \mu_L^\star = \mu_\infty - \frac{ b^2 }{ (\ln L)^2 }.
\end{equation}
In \fig{kt_fit}, we have plotted $\mu_L^\star$ versus $1/(\ln L)^2$, and 
the data are well described by \eq{eq:fit}. From the fit we obtain 
$\mu_\infty = 5.187$, which is also inside the region of the cumulant 
intersections of \fig{u4}. In order to obtain the density $\rho_\infty$ 
at the transition, we have measured
\begin{equation}
 \rho_L^\star \equiv \left. \rho \right|_{\mu = \mu_L^\star},
\end{equation}
as a function of $L$. The result is shown in \fig{rho_kt}, where we have 
assumed that $\rho_L^\star$ is also shifted according to \eq{eq:fit}; by 
fitting we obtain $\rho_\infty \approx 6.98$. Note that \eq{eq:fit} 
probably only approximately describes the density shift, as the latter 
is not a field variable, in contrast to the chemical potential. The fit, 
nevertheless, appears to describe the data well.

\begin{figure}
\begin{center}
\includegraphics[clip=,width=\figwidth]{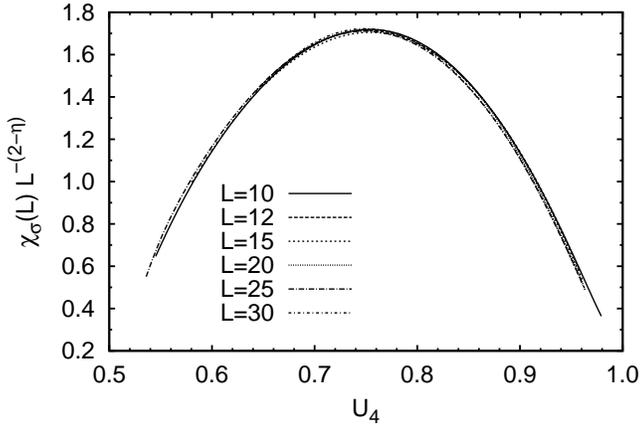}
\caption{\label{loison_sus} Application of Loison's method \cite{loison:1999} to 
obtain the critical exponent $\eta$ in a two-dimensional fluid of hard needles. 
Plotted is $\chi_\sigma(L) \, L^{-(2-\eta)}$ versus $U_4$, for various system 
sizes $L$, using the KT value $\eta=1/4$. The collapse of the data from the 
various system sizes is clearly excellent, and quantitatively confirms the KT 
scenario.}
\end{center}
\end{figure}

\begin{figure}
\begin{center}
\includegraphics[clip=,width=\figwidth]{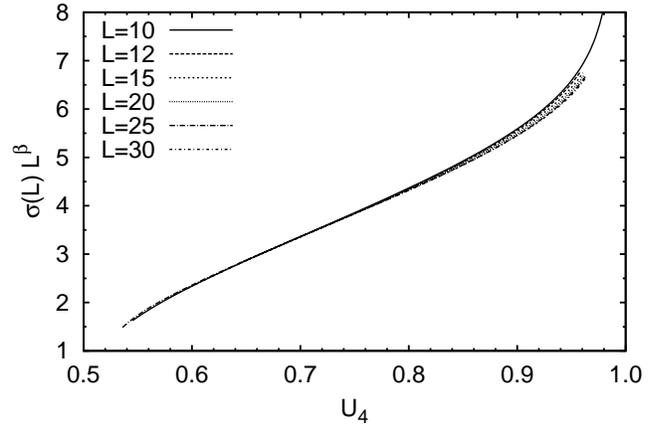}
\caption{\label{loison_op} Application of Loison's method \cite{loison:1999} to 
obtain the critical exponent $\beta$ in a two-dimensional fluid of hard needles. 
Plotted is $\sigma(L) \, L^\beta$ versus $U_4$, for various system sizes $L$, 
using the KT value $\beta=1/8$.}
\end{center}
\end{figure}

We now measure the exponent $\eta$, using the method of Loison 
\cite{loison:1999}. The basic idea is that also the Binder cumulant is 
expressed by a finite size scaling form $U_4 = g (L/\xi)$, with $g$ a 
scaling function. Formally, this can be inverted $L/\xi = g^{-1} (U_4)$; 
substitution into \eq{eq:ansatz} yields
\begin{equation}
 \chi_\sigma (L) \, L^{-(2-\eta)} = h( U_4 ),
\end{equation}
with $h$ another scaling function, which could be expressed in terms of 
$f_2$ and $g$, but the precise form does not matter. Hence, if we plot 
$\chi_\sigma (L) \, L^{-(2-\eta)}$ versus $U_4$, the data from different 
system sizes should collapse onto each other, provided the correct value 
of $\eta$ is used. The result is shown in \fig{loison_sus}, where the KT 
exponent $\eta=1/4$ was used. The collapse of the data from the various 
system sizes is excellent, and gives quantitative confirmation of the KT 
scenario in fluids of hard needles. We observed that the quality of the 
collapse quickly deteriorates when a different exponent $\eta$ is used; 
the numerical uncertainty in $\eta$ is around $\pm 0.01$.

Similarly, for the order parameter, we expect a data collapse when 
$\sigma(L) \, L^\beta$ versus $U_4$ is plotted, provided the correct 
value of $\beta$ is used. The result is shown in \fig{loison_op}, where 
$\beta=1/8$ was used. Again, the collapse is very reasonable, except in 
the \ahum{tails} at high values of $U_4$. Note, however, that here one 
enters the regime $t<0$, where the scaling is expected to break down. 
Compared to the susceptibility, we observed that the quality of the 
collapse is less sensitive to the precise value of $\beta$ being used. 
The numerical uncertainty in $\beta$ is consequently larger, and around 
$\pm 0.05$.

Finally, we discuss the compressibility $\chi_\rho$, see \fig{cv_raw}. The data 
show the formation of a peak, which grows mildly with increasing system size. 
Note also that the maximum occurs well below $\mu_\infty$ of the KT transition. 
Interestingly, compared to lattice simulations of the XY model, $\chi_\rho$ 
behaves conform the specific heat 
\cite{citeulike:4518100,citeulike:4484545,citeulike:4518225,citeulike:4518226}. 
For KT transitions, the exponent of the specific heat is negative, meaning that 
it does not diverge. If we accept that $\chi_\rho$ is the specific heat 
analogue, the most likely scenario is that the peaks in \fig{cv_raw} saturate at 
finite heights in the thermodynamic limit.

\subsection{Conventional critical scaling}

As is well known, there is no magnetization in the XY model in the 
thermodynamic limit. In finite systems, however, the KT transition is 
always accompanied by a rise in magnetization. BH have shown that this 
effect is so strong, it survives in experiments 
\cite{bramwell.holdsworth:1993}. Using renormalization group arguments, 
they demonstrate that the increase in magnetization is conform a 
conventional power law in temperature, with an associated effective 
critical exponent $\beta_{\rm eff} = 3 \pi^2 / 128 \approx 0.23$. More 
remarkably, $\beta_{\rm eff}$ is universal. Indeed, many experiments on 
XY-like systems yield exponents in agreement with the BH prediction 
\cite{citeulike:4472303}. Hence, $\beta_{\rm eff}$ appears to be a 
genuine signature of XY universality, even though in the thermodynamic 
limit it has no meaning.

\begin{figure}
\begin{center}
\includegraphics[clip=,width=\figwidth]{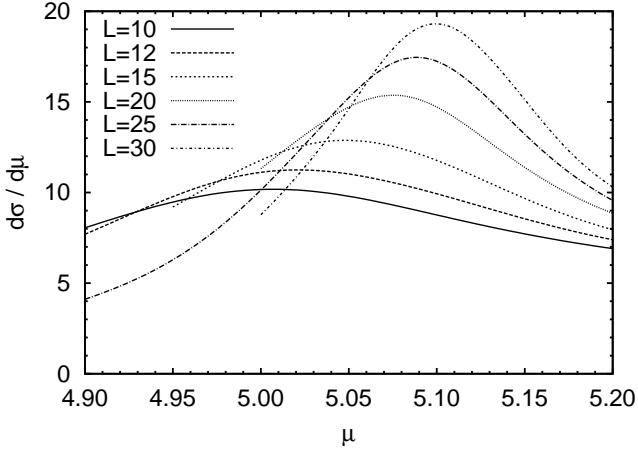}
\caption{\label{slope} Plot of the order parameter \ahum{slope} ${\rm d} \sigma 
/ {\rm d} \mu$ versus the chemical potential $\mu$, for several system sizes $L$ 
as indicated. Note the presence of the maximum, and also that the maximum 
increases with system size. This means that, even though nematic order is absent 
in the thermodynamic limit, finite-sized samples still reveal a steep rise in 
nematic order at special values of the chemical potential.}
\end{center}
\end{figure}

\begin{figure}
\begin{center}
\includegraphics[clip=,width=\figwidth]{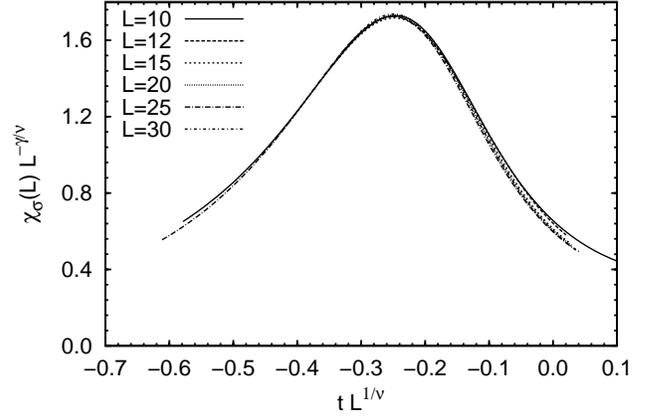}
\caption{\label{sus_conv} Susceptibility scaling plot assuming a 
conventional critical point. Shown is $\chi_\sigma(L) \, 
L^{-\gamma/\nu}$ versus $t L^{1/\nu}$, using $\nu \approx 1.33$, $\gamma 
\approx 2.33$, and $\mu_\infty \approx 5.183$.}
\end{center}
\end{figure}

\begin{figure}
\begin{center}
\includegraphics[clip=,width=\figwidth]{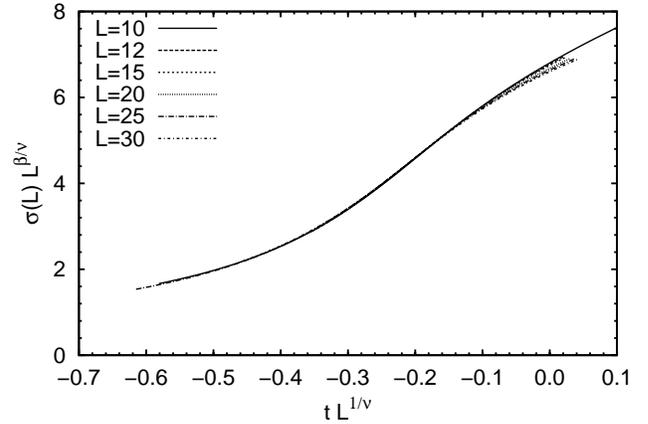}
\caption{\label{op_conv} Order parameter scaling plot assuming a 
conventional critical point. Shown is $\sigma (L) \, L^{\beta/\nu}$ 
versus $t L^{1/\nu}$, using $\nu \approx 1.33$, $\beta \approx 0.18$, 
and $\mu_\infty \approx 5.183$.}
\end{center}
\end{figure}

Considering now the case of two-dimensional hard needles, it seems 
reasonable to expect the validity of the BH scenario also. In the 
vicinity of the KT transition, the order parameter $\sigma$ rises 
sharply, and the slope 
\begin{equation}
 {\rm d} \sigma / {\rm d} \mu = 
 \left( \avg{SN} - \avg{S} \avg{N} \right) / A,
\end{equation}
attains a maximum, see \fig{slope}. Hence, we propose to re-analyze our 
$\sigma$ and $\chi_\sigma$ data, but this time assuming conventional 
critical scaling. The latter is easily done in standard scaling plots 
\cite{newman.barkema:1999}. For the susceptibility, one plots 
$\chi_\sigma(L) \, L^{-\gamma/\nu}$ versus $t L^{1/\nu}$, with $\gamma$ 
and $\nu$ the critical exponents of the susceptibility and correlation 
length, respectively; for the order parameter, one plots $\sigma(L) \, 
L^{\beta/\nu}$ versus $t L^{1/\nu}$, with $\beta$ the critical exponent 
of the order parameter. Recall that $t = \mu_\infty - \mu$ is the 
distance from the critical point. Provided correct values of 
$\mu_\infty$ and the critical exponents are used, the curves from 
different system sizes are expected to collapse.

The results are shown in \fig{sus_conv} and \fig{op_conv}, for the 
susceptibility and order parameter, respectively\footnote{In these plots, the 
relative distance $t = (\mu_\infty - \mu)/\mu_\infty$ was actually used.}. The 
collapse looks reasonable in both cases, and we obtain $\nu \approx 1.33$, 
$\gamma \approx 2.33$, $\beta \approx 0.18$, and $\mu_\infty \approx 5.183$. The 
estimate of $\mu_\infty$ is very close to the KT result of the previous section. 
Obviously, over the range of $L$ available in simulations, deviations from the 
logarithmic shift of \eq{eq:fit} over a power law will be small. Note also that 
the critical exponents obtained from the scaling plots are consistent, in the 
sense that hyperscaling $\gamma + 2 \beta = d \nu$ is obeyed, with $d$ the 
spatial dimension; substitution of our estimates yields $d \approx 2.02$. Hence, 
in agreement with BH, we find that finite systems of needles indeed give rise to 
effective critical exponents. However, our result $\beta_{\rm eff} \approx 0.18$ 
seems rather far removed from the theoretical BH prediction $\beta_{\rm eff,BH} 
\approx 0.23$.

\section{Discussion and Summary}

We have presented grand canonical simulation results of the IN transition of 
hard needles in two spatial dimensions. Our results are consistent with previous 
simulation studies of this model \cite{physreva.31.1776,citeulike:4464987}, and 
confirm that the transition is of the KT type. The novelty of the present work 
has been the combination of multiple histogram reweighting 
\cite{ferrenberg.swendsen:1989} with finite size scaling. This combination 
facilitates an accurate scaling analysis, and critical exponents can be 
meaningfully obtained. Indeed, our data show that the XY exponents $\eta=1/4$ 
and $\beta=1/8$ set in at the transition. The chemical potential and density at 
the transition were found to be $\mu_\infty \approx 5.187$ and $\rho_\infty 
\approx 6.98$, which can be compared to Khandkar and Barma (KB) 
\cite{citeulike:4464987}, who use deposition-evaporation dynamics to study the 
same transition. In the KB case, the control parameter is the ratio of 
deposition-to-evaporation moves $\kappa$; the latter is related to the chemical 
potential via $\kappa = \exp(\mu) / \rho$ \cite{citeulike:4464987}. KB report 
$\kappa_\infty \approx 25.8$, while our estimates of $\mu_\infty$ and 
$\rho_\infty$ yield $\kappa_\infty \approx 25.6$, which is remarkably close.

Considering the behavior of the Binder cumulant, see \fig{u4}, our data reveal 
an approximate intersection point, occurring close to $\mu_\infty$. This 
behavior is conform XY universality 
\cite{loison:1999,citeulike:4541187,citeulike:4541296}. Interestingly, the data 
of KB do not reveal cumulant intersections \cite{citeulike:4464987}. Possibly, 
intersections are also present in the KB data but on a finer $\kappa$ scale, 
accessible only with histogram reweighting 
\cite{ferrenberg.swendsen:1988,ferrenberg.swendsen:1989}. Hasenbusch derived the 
very precise estimate $\lim_{L \to \infty} 1/U_{4,\rm KT} = 1.018192$ for the 
value of the cumulant at a KT transition \cite{citeulike:4402367}; simulation 
data also show that the limit is approached from above with increasing $L$ 
\cite{citeulike:4402367}. Since our definition of the cumulant uses the inverse, 
we anticipate an increase of $U_4$ with increasing $L$. Inspection of \fig{u4} 
reveals a shift of the intersection point toward higher $U_4$ -- consistent with 
Hasenbusch -- but our data still deviate from the theoretical value by $\approx 
3$\%. Presumably, much larger systems are required before the limiting value is 
observed, owing to strong subleading corrections to scaling 
\cite{citeulike:4402367}.

We have also analyzed our data assuming a conventional critical point. The 
motivation was to test whether the effective exponent $\beta_{\rm eff}$, 
predicted by BH, can also be observed. While we do recover effective exponents 
obeying hyperscaling, our estimate of $\beta_{\rm eff}$ does not conform to the 
BH prediction, the deviation being over 20\%. Hence, the most consistent 
description of our results is provided by the KT scenario, in agreement with the 
pioneering simulations of Frenkel and Eppenga \cite{physreva.31.1776}.

\section*{Acknowledgments}

This work was supported by the {\it Deutsche Forschungsgemeinschaft} (DFG) under 
the Emmy Noether program (VI~483/1-1).

\bibliography{mc1975}

\end{document}